\documentclass[12pt]{article}
\usepackage{amssymb,amsmath,amsthm,amsfonts,amscd}
\usepackage{graphicx}
\newcommand\be{\begin{equation}}
\newcommand\ee{\end{equation}}
\newcommand\bea{\begin{eqnarray}}
\newcommand\eea{\end{eqnarray}}

\newcommand{\fatalpha}{{\bf \alpha \kern -0.44em \alpha}}
\newcommand{\fatsigma}{{\bf \sigma \kern -0.54em \sigma}}
\newcommand{\tpchi}{{\bf \chi \kern -0.35em \chi}}
\newcommand{\llambda}{{\bf \lambda \kern -0.45em \lambda}}

\bibliography{plain}
\title{\bf Relation Between Stereographic Projection and Concurrence Measure in Bipartite Pure States}
\author{ G. Najarbashi $^{a}$
 \thanks{Najarbashi@uma.ac.ir} ,B. Seifi $^{a}$ \thanks{B.seifi@uma.ac.ir } \\
\\ \\
$^a${\small Department of Physics, University of Mohaghegh Ardabili, Ardabil 179, Iran.}\\
{\small}} \pagebreak
\begin{document}
\maketitle
\newpage %\vspace{15mm}
\begin{abstract}
One-qubit pure states, living on the surface of Bloch sphere,  can be mapped  onto the usual complex plane by using stereographic projection. In this paper, after reviewing the entanglement of two-qubit pure state,  it is  shown that the \emph{quaternionic} stereographic projection is related to concurrence measure. This is due to the fact that every two-qubit state, in ordinary complex field,  corresponds to the one-qubit state in quaternionic skew field, called quaterbit. Like the one-qubit states in complex field, the stereographic projection  maps every quaterbit onto a quaternion number whose complex and quaternionic parts are related to Schmidt and concurrence terms respectively. Rather,  the same relation is established  for three-qubit state  under  \emph{octonionic} stereographic projection which means that if the state is bi-separable then, quaternionic and octonionic terms vanish. Finally, we  generalize recent consequences to $2 \otimes N$ and $4\otimes N$ dimensional Hilbert spaces ($N\geq 2$) and show that, after stereographic projection,  the quaternionic and octonionic terms are entanglement sensitive.  These trends are easily confirmed by direct computation for general multi-particle  W- and GHZ-states.
\par
 {\bf PACs Index: 03.67.a 03.65.Ud}
\end{abstract}
\newpage
\section{Introduction}
\par
The   Hopf fibration  has a wide variety of physical applications including magnetic monopoles \cite{Nakahar1}, string theory \cite{duf1,levay4,Rios},  new solution of Maxwell equations \cite{ranada}  and  quantum information theory,  \cite{carol,rrd,g.n1,g.n2,g.n3,Nielsen}.
An interesting approach to study the entanglement measure in spin  systems, is geometric structure of multipartite qubit systems \cite{ carol,Zyczkowski,Chruscinski, Schirmer, levay,levay2, levay3,  makela, rau1, rau2,fisc}. Hopf fibration and stereographic projection have simple geometrical interpretation of structure of one, two and three-qubit systems. The relation between stereographic projection, single qubit and two-qubit states has first been studied by Mosseri and Dandoloff \cite{rrd}  in quaternionic projective plane and subsequently has been generalized to three-qubit state based on octonions by Bernevig and Chen \cite{phdm}.
In the study of the dynamical evolutions of a spin-1/2 quantum system, a visual metaphor that has played a powerful role is that of the Bloch sphere. Pure states of the  quantum system are represented by the tip of a unit vector from the origin to the surface of such a unit sphere $S^{2}$. In general, a round $S^{n}$, as a $n$-dimensional sphere, can be embedded  in a flat ($N = n + 1$)-dimensional space (${\mathbb{R}^N}$). Using Cartesian coordinates, the $n$-sphere $S^{n}$ is the surface $\vec{r}.\vec{r}=\sum_{i=0}^{n}x_{i}^2=1$ where $\vec{r}=(x_{0},x_{1},...,x_{n})$. Here, we give a certain standard size (unit radius) to our sphere and also introduce the standard scalar product in ${\mathbb{R}^N}$.
In one-qubit case the stereographic projection maps the point from surface of  Bloch sphere  onto projective complex plane and the single qubit operation on a complex plane is studied in  \cite{Lee}. Two and three-qubit pure states live in surface of unit sphere $S^7$ and $S^{15}$, respectively. As far as computation is concerned, going from the $S^3$ to the$S^7$ and then the $S^{15}$ stereographic projection merely amounts to replacing complex numbers by quaternions and then octonions. The evolution of two and three-qubit can be studied in projective plan, But, there is also one more reason to look for stereographic projection. For two qubit states the concurrence measure \cite{Akhtarshenas} appears explicitly in quaternionic stereographic projection which geometrically means that non-entangled  pure states mapped from  surface of a unit sphere $S^7$ onto a two-dimensional planar subspace of the target space ${\mathbb{R}}^{4}$ \cite{g.n1}. On the other hand it was shown that   octonionic stereographic projection is also entanglement sensitive for three qubit pure states \cite{g.n2,phdm}.
\par
 Quantification of entanglement is an important topic for both theoretical and experimental study. Bennett et al. \cite{Bennett} proposed the concept of the entanglement of formation to quantify entanglement. For a two-qubit pure state the entanglement of formation can be exactly quantified by the concurrence measure introduced by Wootters \cite{Wootters,Hill} which have been generalized for multi-qudit  states in Refs.  \cite{Akhtarshenas,Zhu}. In Ref. \cite{Liu}, the authors describe  an effect method for measuring the concurrence of hyperentanglement by employing the cross-Kerr nonlinearity. There are some other works for concurrence, such as the indirect approach of measuring the atomic entanglement assisted by single photons \cite{Zhou} and  directly measuring the concurrence of a two-qubit electronic pure entangled state by parity-check measurement which is constructed by two polarization beam splitters and a charge detector \cite{Sheng1}.
\par
Consider the $\mathcal{H}_{d}^{\mathbb{C}}$,  $\mathcal{H}_{d}^{\mathbb{Q}}$ and $\mathcal{H}_{d}^{\mathbb{O}}$ for d-dimensional Hilbert space in complex, quaternion and octonion Hilbert space, respectively. It seems that there is also another geometrical approach to describe entanglement between subsystems in $\mathcal{H}_{2}^{\mathbb{C}}\otimes \mathcal{H}_{N}^{\mathbb{C}}$ and $\mathcal{H}_{4}^{\mathbb{C}}\otimes \mathcal{H}_{N}^{\mathbb{C}}$ Hilbert spaces. We show that the quaternionic part of stereographic projection in $\mathcal{H}_{2}^{\mathbb{C}}\otimes \mathcal{H}_{2}^{\mathbb{C}}$ is an entanglement measure that is called concurrence. We generalize this measure to Hilbert space $\mathcal{H}_{2}^{\mathbb{C}}\otimes \mathcal{H}_{N}^{\mathbb{C}}$ also  we show that the octonionic part of stereographic projection in  $\mathcal{H}_{4}^{\mathbb{C}}\otimes \mathcal{H}_{2}^{\mathbb{C}}$ is entanglement sensitive and concurrence between subsystems can be read from octonionic part of stereographic projection and we generalize this  entanglement  measure in $\mathcal{H}_{4}^{\mathbb{C}}\otimes \mathcal{H}_{N}^{\mathbb{C}}$ Hilbert space.
\par
The paper is organized as follows. In section 2, we briefly summarize two-qubit entanglement with use of quaternionic stereographic projection and  we extend  the results of two-qubit systems to systems in Hilbert space $\mathcal{H}_{2}^{\mathbb{C}}\otimes \mathcal{H}_{N}^{\mathbb{C}}$. In section 3, we study the three-qubit  entanglement with the use of octonionic stereographic projection and we extend this results to systems in Hilbert space $\mathcal{H}_{4}^{\mathbb{C}}\otimes\mathcal{H}_{N}^{\mathbb{C}}$.  The paper is ended with a brief conclusion.
\section{Relation between entanglement and stereographic projection in $\mathcal{H}_{2}^{\mathbb{C}}\otimes\mathcal{H}_{N}^{\mathbb{C}}$ Hilbert space }
\subsection{Entanglement in two-qubit pure state ($\mathcal{H}_{2}^{\mathbb{C}}\otimes\mathcal{H}_{2}^{\mathbb{C}}$)}
The general form of two-qubit pure state in Hilbert space ${\mathcal{H}}_{2}^{\mathbb{C}}\otimes{\mathcal{H}}_{2}^{\mathbb{C}}$ is given by
\begin{equation}\label{2qubit}
|\psi\rangle=a_{0}|00\rangle+a_{1}|01\rangle+a_{2}|10\rangle+a_{3}|11\rangle,
\end{equation}
with normalization condition $\sum\limits_{i = 0}^3 {|{a_i}{|^2}}=1$ and the  direct product basis $\{|00\rangle,|01\rangle,|10\rangle,|11\rangle\}$. The two-qubit state  $|\psi\rangle$ is said separable if it can be written as a simple product of individual kets in separated Hilbert space $\mathcal{H}_{2}^{\mathbb{C}}$, a definition which translates into the well known
 condition $a_{1}a_{2}=a_{0}a_{3}$. The generic state (\ref{2qubit}) is not separable and is said to be entangled. The  normalization condition, identifies $|\psi\rangle$ to the 7-dimensional sphere $S^7$, embedded in $\mathbb{R}^8$. Therefore this tempting to see whether the known $S^7$ Hopf fibration (with fibres $S^3$ and base $S^4$) can play any role in the Hilbert space description. To introduce the stereographic projection for two-qubit  we first must represent the state (\ref{2qubit}) in quaternionic representation as
\begin{equation}\label{quaterbit}
\mathcal{Q}|\psi\rangle=|\psi\rangle_{q}:=q_{0}|0\rangle_{q}+q_{1}|1\rangle_{q},
\end{equation}
where $q_{0}=a_{0}+a_{1}e_{2}$ and $q_{1}=a_{2}+a_{3}e_{2}$ are two quaternion numbers. Quaternion is an associative and noncommutative algebra of
rank 4 on real space $\mathbb{R}$ whose every element can be written as $q=\sum\limits_{i = 0}^3 {{x_i}{e_i}},\ {x_i} \in \mathbb{R}$
where $e_{0}=1$ and the product of two quaternion can be calculated according to Table \ref{table1}. The  quaternion can be equivalently defined in term of complex numbers $z_{1}=x_{0}+x_{1}e_{1}$  and  $z_{2}=x_{2}+x_{3}e_{1} $ in the form  $q=z_{1}+z_{2}e_{2} $ with an involutory anti-automorphism (conjugation) such as $\bar{q}=x_{0}-\sum\limits_{i = 1}^3 {{x_i}{e_i}}=\bar{z}_{1}-z_{2}e_{2} $.
 Note that in term of quaternion numbers the normalization condition of state (\ref{quaterbit}) is given by $|q_{0}|^{2}+|q_{1}|^{2}=1$. In quaternion language the ket (\ref{2qubit}) can be restate as (\ref{quaterbit}) with the following representation
\begin{table}[table1]
\centering
\begin{tabular}{|c|| c| c| c| c|}
 \hline
$\times$ &$e_0$&$e_1$&$e_2$&$e_3$ \\ [0.8ex]
 \hline\hline
$e_0$&1&$e_1$&$e_2$&$e_3$ \\  \hline
$e_1$&$e_1$&-1&$e_3$&$-e_2$ \\  \hline
$e_2$&$e_2$&$-e_3$&-1&$e_1$  \\  \hline
$e_3$&$e_3$&$e_2$&$-e_1$&-1  \\   \hline
\end{tabular}
\caption{ Multiplication table of the unit quaternions}
\label{table1}
\end{table}
\begin{align}\label{q-basis}
\begin{array}{c}
|00\rangle\ \ \ \  \longrightarrow \ \ \ \ \ \ |0\rangle_{q},\\
|01\rangle\ \ \ \ \longrightarrow \ \ \ \ e_{2}|0\rangle_{q},\\
|10\rangle\ \ \ \  \longrightarrow \ \ \ \ \ \ |1\rangle_{q},\\
|11\rangle\ \ \ \  \longrightarrow \ \ \ \   e_{2}|1\rangle_{q},
\end{array}
\end{align}
The Hopf fibration (${S^7}\xrightarrow{{{S^3}}}{\mathbb{R}^4}$) from  total space $S^7$ to  base space $\mathbb{R}^4$ with fiber space $S^3$ is defined by  the Hopf map of quaterbit (\ref{quaterbit}), as \cite{rrd,phdm}
\begin{equation}\label{1}
\mathcal{P}|\psi\rangle_{q}:=q_{0}\bar{q}_{1}= S+Ce_{2},
\end{equation}
where $S=a_{0}\bar{a}_{2}+a_{1}\bar{a}_{3}$ and  $C=a_{0}a_{3}-a_{1}a_{2}$ denote respectively the Schmidt and concurrence terms. For $S=0$  the two-qubit pure state have  Schmidt form  $\sqrt{\lambda}|00\rangle+\sqrt{1-\lambda}|11\rangle$ . On the other hand, for two-qubit pure state   $\mathcal{C}=2|C|$ is concurrence measure and for $\mathcal{C}=0$ the two qubit pure state (\ref{2qubit}) is reduced to a separable state, i.e.  factorized as $|\psi\rangle_{AB}=|\psi\rangle_{A}\otimes|\psi\rangle_{B}$. We can show that the entanglement term $C$ in Eq. (\ref{1}) is invariant under local unitary transformation in both complex representation (\ref{2qubit}) and quaternion representation (\ref{quaterbit}).
To this aim we show that the following diagram is commutative
\[\begin{CD}
{\mathcal{H}}_{4}^{\mathbb{C}}         @>{\mathcal{Q}}>>      {\mathcal{H}}_{2}^{\mathbb{Q}}     \\
@V{B}VV               @V{\mathbb{Q}B}VV           \\
{\mathcal{H}}_{4}^{\mathbb{C}}       @>{\mathcal{Q}}>>
{\mathcal{H}}_{2}^{\mathbb{Q}}
\end{CD}\]
where $\mathcal{Q}$ denotes the quaternification operation which is defined in Eq. (\ref{quaterbit}), and $B=A \otimes A'\in SU(2)\otimes SU(2)$ is any local unitary transformation acting on two-qubit pure state (\ref{2qubit})
\begin{equation}\label{trans1}
 |\psi\rangle'=A \otimes A'|\psi\rangle, \ \ \  \ A, A'\in SU(2).
\end{equation}
Note that, $SU(2)$ transformation can be parameterized in the complex field as
\begin{equation}
A=\left( {\begin{array}{*{20}{c}} a& b \\  { - \bar b}& \bar{a}
\end{array}} \right),\quad\quad |a|^2+|b|^2=1.
\end{equation}
We may alternatively write the transformation (\ref{trans1}) in quaternionic   right module as follows
\begin{equation}\label{module}
|{\psi}\rangle' _{q}=\mathbb{Q}B(|{\psi}\rangle_{q}):= A | {\psi}\rangle_{q}{\mathcal{A}'}^{(q)}= \left( {\begin{array}{*{20}{c}} a& b \\  { -\bar b}&\bar{a}
\end{array}} \right)\left(\begin{array}{c}
                      q_{0} \\
                      q_{1}
                    \end{array}\right)(a'-\bar b' e_{2})=\left(\begin{array}{c}
                      q'_{0} \\
                      q'_{1}
                    \end{array}\right),
\end{equation}
where $q'_{0}=aq_{0}(a'-\bar b' e_{2})+bq_{1}(a'-\bar b' e_{2})$, $q'_{1}=-\bar{b}q_{0}(a'-\bar b' e_{2})+\bar{a}q_{1}(a'-\bar b' e_{2})$
and  the definition  ${\mathcal{A}'}^{(q)}=a'-\bar b' e_{2}$ as quaternionic representation for unitary matrix $A'$ come from the isomorphism $Sp(1)\simeq SU(2)$. The second equality by itself defines the $\mathbb{Q}B$ operation in the diagram.
The stereographic projection $|{\psi}\rangle' _{q}$ according to $q'_{0}$ and $q'_{1}$ is
\begin{equation}\label{2}
\mathcal{P}|\psi\rangle'_{q}:=q'_{0}\bar{q}'_{1}= S'+C'e_{2},
\end{equation}
where $C'=C=(\beta\gamma-\alpha\delta)$ is  concurrence term which  remains
invariant under the transformation $|\psi\rangle\to|\psi\rangle'$, while
\begin{equation}
S'=(|q_{1}|^{2}-|q_{0}|^{2})ab+S|a|^{2}-\bar S |b|^{2},
\end{equation}
is  not invariant Schmidt term under local unitary groups. Summarizing, we have
\begin{equation}\label{ww}
\mathcal{P}\mathcal{Q}B|\psi\rangle
=\mathcal{P}(\mathbb{Q}B)\mathcal{Q}|\psi\rangle.
\end{equation}
It should be mentioned that  $\mathbb{Q}B$ acting on quaterbit $\mathcal{Q}|\psi\rangle$, endowed with right module as Eq. (\ref{module}), belongs to the general form of local unitary transformation   implying that the quaternionic part of stereographic projection (\ref{1}) is also an entanglement measure for two-qubit pure states.
\par
 In quantum mechanics, according  to Schr\"{o}dinger equation
$
i\hbar \frac{\partial }{{\partial t}}|\psi(t) \rangle = {\hat H}|\psi(t) \rangle ,
$
all  dynamical information about any quantum mechanical system is contained in the matrix elements of its time evolution operator
\begin{equation}
|\psi(t) \rangle  =e^{- i{\hat H }t}|\psi(0) \rangle,
\end{equation}
where $e^{- i{\hat H} t}\equiv A\in SU(2)$ describes dynamical evolution, up to a global phase, under the influence of the Hamiltonian from time zero to time t.
For two-qubit quantum systems with initial state $\sqrt{\lambda}|00\rangle+\sqrt{1-\lambda}|11\rangle$, we have
\begin{equation}
|\psi(t)\rangle  =e^{- i{\hat{H} }_{1}t}\otimes e^{- i{\hat{H} }_{2}t}(\sqrt{\lambda}|00\rangle+\sqrt{1-\lambda}|11\rangle),
\end{equation}
with local Hamiltonians
\begin{equation}\label{hamiltonian}
H_{i} =\vec r_{i}.\vec \sigma ,\quad \vec r_{i}=r\sin\theta_{i}\cos\phi_{i}\hat{i}+r\sin\theta_{i}\sin\phi_{i}\hat{j}+r\cos\theta_{i}\hat{k},
\end{equation}
where $\sigma_{i}$ are  usual pauli matrices and $\hat{i}, \hat{j}$ and $\hat{k}$ are basis of Euclidean space. The quaternionic form  of state $|\psi(t)\rangle$ is
\begin{equation}
|\psi(t)\rangle_{q}=\mathcal{Q}|\psi(t)\rangle=q_{0}|0\rangle_{q}+q_{1}|1\rangle_{q},
\end{equation}
with
\begin{align}\label{ooo}
\begin{array}{c}
q_{0}=[\sqrt{\lambda } \left(\cos \left(\frac{t}{2}\right)-i \cos \left(\theta _1\right) \sin \left(\frac{t}{2}\right)\right) \left(\cos \left(\frac{t}{2}\right)-i \cos \left(\theta _2\right) \sin \left(\frac{t}{2}\right)\right) \quad \quad \quad \quad \quad \quad \quad \quad \\
-\sqrt{1-\lambda } e^{-i\left(\phi _1+\phi _2\right)} \sin \left(\theta _1\right) \sin \left(\theta _2\right) \sin ^2\left(\frac{t}{2}\right) ]\quad \quad \quad \quad\quad \quad \quad\quad \quad \quad \quad \quad \quad   \\
+[-i\sin \left(\frac{t}{2}\right)\left(\sqrt{1-\lambda } \sin \left(\theta _1\right) e^{i\phi_2} \left(\cos \left(\frac{t}{2}\right)+i \cos \left(\theta _2\right) \sin \left(\frac{t}{2}\right)\right)\right.\quad \quad \quad \quad \quad \quad\quad   \\
\left.+\sqrt{\lambda } \sin \left(\theta _2\right)e^{i\phi_2} \left(\cos \left(\frac{t}{2}\right)-i \cos \left(\theta _1\right) \sin \left(\frac{t}{2}\right)\right)\right)]e_{2},  \quad \quad \quad \quad \quad \quad \quad \quad \quad \quad \\
q_{1}=[-i\sin \left(\frac{t}{2}\right)\left(\sqrt{\lambda } \sin \left(\theta _1\right) e^{i\phi_1} \left(\cos \left(\frac{t}{2}\right)-i \cos \left(\theta _2\right) \sin \left(\frac{t}{2}\right)\right)\right.  \quad \quad \quad\quad \quad \quad \quad \quad \quad\quad \\
\left.+\sqrt{1-\lambda } \sin \left(\theta _2\right) e^{-i\phi_2} \left(\cos \left(\frac{t}{2}\right)+i \cos \left(\theta _1\right) \sin \left(\frac{t}{2}\right)\right)\right) ]\quad \quad \quad \quad \quad \quad \quad \quad \quad  \\
+[\sqrt{1-\lambda } \left(\cos \left(\frac{t}{2}\right)+i \cos \left(\theta _1\right) \sin \left(\frac{t}{2}\right)\right) \left(\cos \left(\frac{t}{2}\right)+i \cos \left(\theta _2\right) \sin \left(\frac{t}{2}\right)\right)\quad \quad \quad \quad \quad  \\
-\sqrt{\lambda } \sin \left(\theta _1\right) \sin \left(\theta _2\right) \sin ^2\left(\frac{t}{2}\right)e^{i\left(\phi _1+\phi _2\right)} ]e_{2},\quad \quad \quad \quad\quad \quad \quad \quad \quad \quad \quad \quad \quad \quad
\end{array}
\end{align}
\begin{figure}
  \centering
  \includegraphics[width=14 cm]{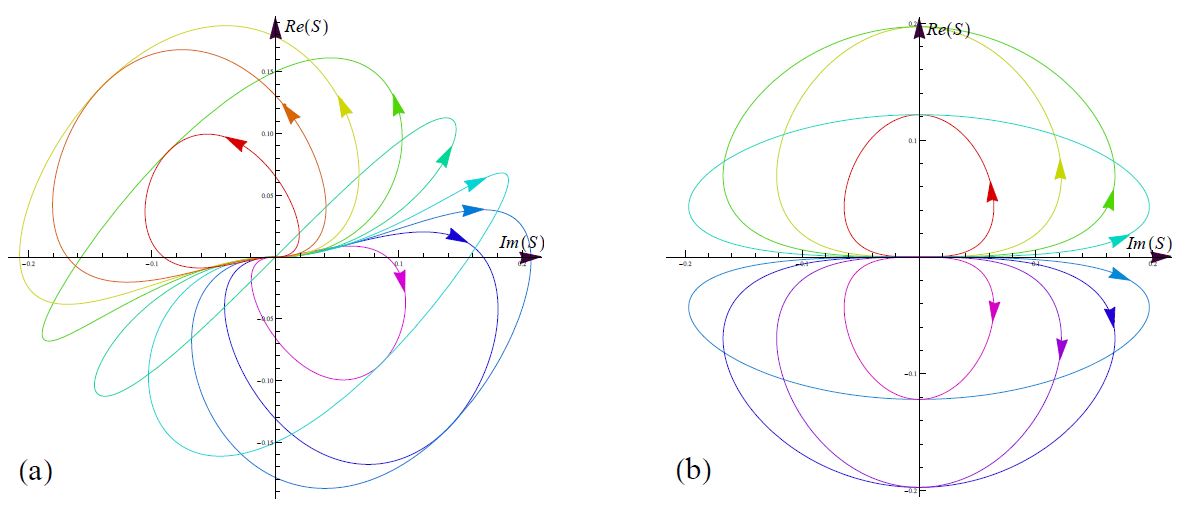}\\
  \caption{\small { Time evolution of Schmidt term for initial states $|\psi\rangle=\frac{1}{\sqrt{2}}(|00\rangle+|11\rangle)$ with $\phi_{1}=0$ (for $a$), $\phi_{1}=\frac{\pi}{2}$ (for $b$) and $\theta_{1}=\frac{\pi}{10}, \frac{2\pi}{10}, ..., \pi$  in Hamiltonian (\ref{hamiltonian}).}}\label{appendix Fig.1}
\end{figure}
and stereographic projection of state $|\psi(t)\rangle_{q}$ is given by
\begin{equation}
\mathcal{P}|\psi(t)\rangle_{q}=S'+C' e_{2},
\end{equation}
where $C'=\sqrt{\lambda(1-\lambda)}$ is invariant concurrence term and $S'=Re(S')+Im(S')e_{2}$ is Schmidt term with real and imaginary parts:
\begin{align}\label{shmidt}
\begin{array}{c}
 Re(S')=(2 \lambda -1) \sin \left(\frac{t }{2}\right) \sin \left(\theta _1\right) \left(\sin \left(\frac{t }{2}\right) \cos \left(\theta _1\right) \cos \left(\phi _1\right)+\cos \left(\frac{t}{2}\right) \sin \left(\phi _1\right)\right) ,\\
  Im(S')=(2 \lambda -1) \sin \left(\frac{t }{2}\right) \sin \left(\theta _1\right) \left(\cos \left(\frac{t}{2}\right) \cos \left(\phi _1\right)-\sin \left(\frac{t }{2}\right) \cos \left(\theta _1\right) \sin \left(\phi _1\right)\right),
\end{array}
\end{align}
(see the Fig. (\ref{appendix Fig.1}), where the time evolution of Schmidt term is sketched in complex plane.)  Notice that the parameters of $\hat{H}_{2}$ do not appear in concurrence and Schmidt term of stereographic projection. This example shows  that the evolution of  second qubit is established in the fiber space of Hopf fibration.
\subsection{Entanglement in $\mathcal{H}_{2}^{\mathbb{C}}\otimes\mathcal{H}_{N}^{\mathbb{C}}$ Hilbert space}
In this section, we  generalize the relation between quaternionic stereographic projection and  concurrence  measure for a  bipartite pure state in Hilbert space $\mathcal{H}_{2}^{\mathbb{C}}\otimes\mathcal{H}_{N}^{\mathbb{C}}$ . For this purpose, let us first consider a pure state   $|\psi\rangle\in\mathcal{H}_{2}^{\mathbb{C}}\otimes\mathcal{H}_{N}^{\mathbb{C}}$  in  the following
generic form
\begin{equation}\label{2part}
|\psi\rangle=\sum\limits_{i = 0}^1 {\sum\limits_{j = 0}^{N-1} {{a_{ij}}\left| {ij} \right\rangle } } ,
\end{equation}
where $|i\rangle (i=0,1)$  and  $|j\rangle (j=0,1,...,N-1)$ are orthonormal  basis of Hilbert space $\mathcal{H}_{2}^{\mathbb{C}}$ and $\mathcal{H}_{N}^{\mathbb{C}}$ respectively. The concurrence of state (\ref{2part}) is \cite{Akhtarshenas}
 \begin{equation}\label{concurrence1}
C=\sqrt {2\sum\limits_{j \neq k=0}^{N-1} {|{a_{0j}}{a_{1k}} - {a_{1j}}{a_{0k}}{|^2}} }=\sqrt {\sum\limits_{\alpha  = 0}^{\frac{{N(N - 1)}}{2}} {|\langle \psi |\tilde \psi_{\alpha} \rangle {|^2}} },
\end{equation}
where  $|\tilde \psi _{\alpha}\rangle=(S\otimes L_{\alpha})|\psi^*\rangle$, $S=i\sigma_{2}$ is the generator of   $SO(2)$ and $L_{\alpha}$ are generators of $SO(N)$ group with matrix elements $ (L_{\alpha})_{kl} = (L_{[j_0j_1...j_{N-3}]})_{kl} =\varepsilon_{[j_0j_1...j_{n-3}]kl}$ and $\varepsilon_{j_0j_1....j_{N-1}}$ is totally antisymmetric under interchange of any two indices with $\varepsilon_{01...N-1} =1$. For two-qubit case we show that the quaternionic part of stereographic projection  $C$ is an entanglement measure. It is interesting to generalize the quaternification operation  and  stereographic projection to state (\ref{2part}) and investigate the concurrence term in the language of quaternionic stereographic projection. Just like two-qubit case we can define the quternionic form of state (\ref{2part}) as
\begin{equation}\label{2qpart}
|\psi\rangle_{q}= {\sum\limits_{j =0}^{N-1} {{q_{j}}\left| j \right\rangle_{q} } } ,
\end{equation}
where  $| j \rangle_{q}$ is defined by
\begin{align}\label{qua1}
\begin{array}{c}
|0\alpha\rangle \ \ \  \longrightarrow \ \ \ \ \ \ |\alpha\rangle_{q}, \\
|1\alpha\rangle \ \ \  \longrightarrow \ \ \ e_{2}|\alpha\rangle_{q}.
\end{array}
\end{align}
 The quaternion stereographic projection between $q_{j}$ and   $q_{k}$  of quaternionic state (\ref{2qpart}) is defined  as $q_{i}\bar{q}_{j}=S_{j,k}+C_{j,k}e_{2}$ where  $S_{j,k}=a_{0j}\bar{a}_{0k}+a_{1j}\bar{a}_{1k}$ and  $C_{j,k}=a_{0j}a_{1k} - a_{1j}a_{0k}$ are Schmidt and concurrence terms respectively. Then  the concurrence (\ref{concurrence1}) can be rewritten in the following form
\begin{equation}\label{c1}
C=\sqrt{2\sum\limits_{i \ne j=0}^{N-1} {|{C_{i,j}}{|^2}}}.
\end{equation}
This means that the concurrence of quaternionic state  (\ref{2qpart}) is given  by  the norm of concurrence vector with element $\sqrt{2}C_{i,j}$. For example in  three-qubit pure state
\begin{align}\label{multi1}
\begin{array}{c}
|\psi\rangle=z_{000}|000\rangle+z_{001}|001\rangle+z_{010}|010\rangle+z_{011}|011\rangle \\
\ \ \  \ \ \ \ \ z_{100}|100\rangle+z_{101}|101\rangle+z_{110}|110\rangle+z_{111}|111\rangle,
\end{array}
\end{align}
the quaternionic form  is given by
\begin{equation}\label{q1}
\mathcal{Q}|\psi\rangle=|\psi\rangle_{q}=q_{00}|00\rangle_{q}+q_{01}|01\rangle_{q}+q_{10}|10\rangle_{q}+q_{11}|11\rangle_{q},
\end{equation}
where $q_{ij}=z_{0ij}+z_{1ij}e_{2}$ and the concurrence of this state is given by
\begin{equation}\label{qp}
C=2\sqrt{|C_{11,00}|^{2}+|C_{11,01}|^{2} +|C_{11,10}|^{2}+|C_{10,00}|^{2}+|C_{10,01}|^{2}+|C_{01,00}|^{2}},
\end{equation}
where $C_{ij,kl}$ is quaternionic part of stereographic projection with respect to $q_{ij}$ and  $q_{kl}$.
\par
Let us consider two special cases. The first is the general GHZ-state
\begin{equation}\label{ghz}
|GHZ_{m}\rangle=(|0^{\otimes m}\rangle+|1^{\otimes m}\rangle)/\sqrt{2}.
\end{equation}
Quaternion form of this state is given by $|GHZ_{m}\rangle_{q}=(|0^{\otimes m-1}\rangle_{q}+e_{2}|1^{\otimes m-1}\rangle_{q})/\sqrt{2}$ ($q_{0}=1/\sqrt{2}$ and $q_{1}=e_{2}/\sqrt{2}$). For this state  Eq. (\ref{qp}) give $C^{1, \{m-1\}} =1$.
\par
As another example, let us consider the general W-state defined as
\begin{equation}\label{w}
|W_{m}\rangle=(|000...001\rangle+|000...010\rangle+...+|010...000\rangle+|100...000\rangle)/\sqrt{m}.
\end{equation}
Quaternion form of this state is given by
 $|W_{m}\rangle_{q}=(|00...001\rangle_{q}+|00...010\rangle_{q}+...+|01...000\rangle_{q}+|10...000\rangle_{q})/\sqrt{m}+e_{2}|00...000\rangle_{q}/\sqrt{m}$
 ($q_{0}=...=q_{m-2}=1/\sqrt{m}$ and $q_{m-1}=e_{2}/\sqrt{m}$). For these states we obtain $C^{1,\{m-1\}}=\frac{2\sqrt{m-1}}{m}$.
\section{Relation between entanglement and stereographic projection in $\mathcal{H}_{4}^{\mathbb{C}}\otimes\mathcal{H}_{N}^{\mathbb{C}}$ Hilbert space }
\subsection{Entanglement  in three-qubit pure state ($\mathcal{H}_{4}^{\mathbb{C}}\otimes\mathcal{H}_{2}^{\mathbb{C}}$ case)}
The general form of three-qubit pure state in Hilbert space ${\mathcal{H}}_{2}^{\mathbb{C}}\otimes{\mathcal{H}}_{2}^{\mathbb{C}}\otimes{\mathcal{H}}_{2}^{\mathbb{C}}$ is given by
\begin{equation}\label{threequbit}
|\psi\rangle=t_{0}|000\rangle+t_{1}|001\rangle+
t_{2}|010\rangle+t_{3}|011\rangle+t_{4}|100\rangle+t_{5}|101\rangle+
t_{6}|110\rangle+t_{7}|111\rangle ,
\end{equation}
where $\{|000\rangle, ....,|111\rangle\}$ are basis of Hilbert space and  $t_{i}\in \mathbb{C}$ are complex numbers that satisfy the normalization condition
$\sum_{i=0}^7|t_{i}|^2=1$.
We can equivalently rewrite every $|\psi\rangle\in {\mathcal{H}}_{8}^{\mathbb{C}}$ by two-quaterbit $|{\psi}\rangle_{q}\in {\mathcal{H}}_{4} ^{\mathbb{Q}}$ with   projecting onto a basis  which all coefficients are singly quaternionic number as
\begin{equation}\label{quater1}
\mathcal{Q}(|\psi\rangle):=|{\psi}\rangle_{q}=q_{0}|{00}\rangle_{q}+q_{1}|{01}\rangle_{q} +q_{2}|{10}\rangle_{q} +q_{3}|{11}\rangle_{q},
\end{equation}
where  $|i j \rangle_{q}$ is defined by
\begin{align}\label{qua11}
\begin{array}{c}
|0ij\rangle \ \ \  \longrightarrow \ \ \ \ \ \ |ij\rangle_{q}, \\
|1ij\rangle \ \ \  \longrightarrow \ \ \ e_{2}|ij\rangle_{q},
\end{array}
\end{align}
and $q_{0}, ..., q_{3}$ are quaternion numbers which is given by
\begin{equation}
q_{0}=t_{0}+t_{2}e_{2}\quad,q_{1}=t_{4}+t_{6}e_{2}\quad,q_{2}=t_{1}+t_{3} e_{2}\quad,q_{3}=t_{5}+t_{7}e_{2},\quad
\end{equation}
with normalization condition $\sum_{i=0}^3|q_{i}|^2=1$. Once again, we can equivalently rewrite every $|\psi\rangle_{q} \in {\mathcal{H}}_{4}^{\mathbb{Q}}$ as an octobit $|{\psi}\rangle_{o}\in {\mathcal{H}}_{2} ^{\mathbb{O}}$
 with  the following representation
\begin{align}\label{qua12}
\begin{array}{c}
q_{0}|00\rangle_{q} \ \ \  \longrightarrow \ \ \ \ \ \ q_{0}|0\rangle_{o}, \\
q_{1}|01\rangle_{q} \ \ \  \longrightarrow \ \ \ \ q^{*}_{1}e_{4}|0\rangle_{o}, \\
q_{2}|10\rangle_{q} \ \ \  \longrightarrow \ \ \ \ \ \ q_{2}|1\rangle_{o}, \\
q_{3}|11\rangle_{q} \ \ \  \longrightarrow \ \ \ \  q^{*}_{3}e_{4}|1\rangle_{o},
\end{array}
\end{align}
where $q^{*}_{1}=t_{4}-t_{6}e_{2}$ and  $q^{*}_{3}=t_{5}-t_{7}e_{2}$. Then the octonionic form of state (\ref{threequbit}) can be written as
\begin{equation}\label{octabit}
|{ \psi }\rangle_{o}=\mathcal{O}(\mathcal{Q}(|\psi\rangle))=\mathcal{O}|{\psi}\rangle_{q}=o_{0}|{0 }\rangle_{o}+o_{1}|{ 1 }\rangle_{o},
\end{equation}
where $o_{0}=q_{0}+q^{*}_{1}e_{4}$ and $o_{1}=q_{2}+q^{*}_{3}e_{4}$ are octonion numbers.
\begin{table}[table2]
\centering
\begin{tabular}{|c|| c| c| c| c| c| c| c| c|}
 \hline
$\times$ &$e_0$&$e_1$&$e_2$&$e_3$ &$e_4$&$e_5$&$e_6$&$e_7$ \\ [0.8ex]
 \hline\hline
$e_0$&$1$&$e_1$&$e_2$&$e_3$&$e_4$&$e_5$&$e_6$&$e_7$ \\  \hline
$e_1$&$e_1$&$-1$&$e_3$&$-e_2$&$e_5$&$-e_4$&$-e_7$&$e_6$ \\  \hline
$e_2$&$e_2$&$-e_3$&$-1$&$e_1$&$e_6$&$e_7$&$-e_4$&$-e_5$  \\  \hline
$e_3$&$e_3$&$e_2$&$-e_1$&$-1$&$e_7$&$-e_6$&$e_5$&$-e_4$  \\   \hline
$e_4$&$e_4$&$-e_5$&$-e_6$&$-e_7$&$-1$&$e_1$&$e_2$&$e_3$  \\   \hline
$e_5$&$e_5$&$e_4$&$-e_7$&$e_6$&$-e_1$&$-1$&$-e_3$&$e_2$  \\   \hline
$e_6$&$e_6$&$e_7$&$e_4$&$-e_5$&$-e_2$&$e_3$&$-1$&$-e_1$  \\   \hline
$e_7$&$e_7$&$-e_6$&$e_5$&$e_4$&$-e_3$&$-e_2$&$e_1$&$-1$  \\   \hline
\end{tabular}
\caption{ Multiplication table of the unit octonions}
\label{table2}
\end{table}
\par
The octonionic numbers $\mathbb{O}$ form a non-commutative and non-associative algebra of
rank 8 over $\mathbb{R}$ whose every element can be written as
\begin{equation}\label{octan}
o=\sum\limits_{i = 0}^7 {{x_i}{e_i}} \quad \quad ,{x_i} \in \mathbb{R}\quad \quad ,{e_0} = 1\quad \mathrm{and}  \quad e_i^2 =  - 1{\kern 1pt} {\kern 1pt} (i = 1,...,7).
\end{equation}
The multiplication of octonions is given by ${e_i}{e_j} =  - {\delta _{ij}}{e_0} +\sum\limits_{k = 1}^7 {{f_{ijk}}{e_k}} ,( i,j = 1,2,...,7)$
where ${f_{ijk}}$, just as  Levi-Civita symbol, are totally anti-symmetric in $i, j$ and $k$ with values $1, 0,-1$ and, $f_{ijk} = +1$ for $ijk= 123, 145, 246, 347,617, 725,536$.
Multiplication table of the unit octonions (Table \ref{table2}) implies that the octonion numbers is non-associative and one must keep the parenthesis in all equations involving the product of three or more octonions   i.e. $(o_{1}o_{2})o_{3}\neq o_{1}(o_{2}o_{3})$.
 The complex conjugate of  octonion (\ref{octan}) is given by
\begin{equation}\label{bar}
\bar o=x_{0}-\sum\limits_{i = 1}^7 {{x_i}{e_i}}.
\end{equation}
Equivalently, each octonion can be represented  in terms of four complex number $ z_{0}=x_{0}+x_{1}e_{1}$, $ z_{1}=x_{2}+x_{3}e_{1}$, $z_{2}=x_{4}+x_{5}e_{1}$ and $z_{3}=x_{6}+x_{7}e_{1}$ as $o=z_{0}+z_{1}e_{2}+(z_{3}+z_{3}e_{2})e_{4}$
and complex conjugate of this octonion  is given by $\bar o=\bar{z}_{0}- z_{1}e_{2}-(z_{3}+z_{3}e_{2})e_{4}$. Every non-zero ($o\in \mathbb{O}$)  is invertible, and the unique inverse is given by $o^{-1}=\frac{1}{|o|^2}\bar{o}$ where the octonionic  norm $|o|$ is defined by $ |o|^2=o\bar{o}$. The norm of two octonions  $o_{1}$ and $o_{2}$ satisfies $|o_{1}o_{2}| = |o_{2}o_{1}| = |o_{1}||o_{2}|$.
\par
Clearly, $|{ \psi }\rangle_{o}$ in Eq. (\ref{octabit}) is an octobit (octonion bit) which  belongs to the two dimensional octonionic Hilbert space ${\mathcal{H}}_{2}^{\mathbb{O}}$.
Regarding the above considerations, the octonionic stereographic projection $\mathcal{P}$ is defined as
\begin{equation}\label{confo2}
\mathcal{P}(|{\psi}\rangle_{o}):= o_{0} \bar{o}_{1}=(S_{0}+S_{1}e_{2}+(S_{2}+S_{3}e_{2})e_{4})\in \mathbb{O},
\end{equation}
where
\begin{align}\label{conf}
\begin{array}{c}
  {S_0} =t_{0}\bar{t_{1}}+ t_{2}\bar{t_{3}}+ t_{4}\bar{t_{5  }}+ t_{6}\bar{t_{7}},\\
  {S_1} =t_{2  }t_{1  }- t_{0}t_{3  }+\bar{t_{6  }}\bar{t_{5  }}-\bar{t_{4  }}\bar{t_{7}},\\
  {S_2} =  t_{4  }t_{1} - t_{0}t_{5}+\bar{t_{2  }}\bar{t_{7}}-\bar{t_{6  }}\bar{t_{3}},\\
  {S_3} = t_{4  }t_{3  } - t_{2  }t_{5  }+\bar{t_{6  }}\bar{t_{1  }}-\bar{t_{0}}\bar{t_{7}}.
\end{array}
\end{align}
The $S_{1}, S_{2}$ and $S_{3}$  parts of  stereographic projection are entanglement sensitive for three-qubit  pure state, namely, for separable three-qubit state $|\psi\rangle_{ABC}=|\psi\rangle_{AB}|\psi\rangle_{C}$ the $C=S_{1}e_{2}+(S_{2}+S_{3}e_{2})e_{4}$  vanishes.

In doing so, we will express the concurrence of three-qubit state (\ref{threequbit}) in terms of stereographic projection. To this end we assume the general form of bipartite pure state
\begin{equation}\label{q}
\left| \psi  \right\rangle_{AB}  = \sum\limits_{i = 0}^{N_1-1 } {\sum\limits_{j = 0}^{N_2-1 } {a_{ij} \left| {i \rangle \otimes |j } \right\rangle }},
\end{equation}
 which has concurrence
\begin{equation}\label{cc}
C^{AB} = 2  \sqrt {\sum\limits_{i < j}^{N_1 -1} {\sum\limits_{k < l}^{N_2-1 } {\left| {a_{ik} a_{jl} - a_{il} a_{jk} } \right|^2 } } }.
\end{equation}
According to state (\ref{threequbit}), if we  take the first two-qubit as partition $A$ and the last qubit as partition $B$, then its concurrence can be calculated in terms of coefficients $t_{i}$ as
\begin{equation}\label{1(23)}
\begin{array}{l}
C^{(12)3}=2(|t_{0}t_{3}-t_{1}t_{2}|^{2}+|t_{0}t_{5}-t_{1}t_{4}|^{2}+|t_{0}t_{7}-t_{1}t_{6}|^{2}\\
\quad\quad \quad\quad +|t_{2}t_{5}-t_{3}t_{4}|^{2}+|t_{2}t_{7}-t_{3}t_{6}|^{2}+|t_{4}t_{7}-t_{5}t_{6}|^{2})^{\frac{1}{2}}.\\
 \end{array}
\end{equation}
Now if $C^{(12)3}=0$, i.e. last qubit is factorized from the first two qubits, then the terms  $S_{1},S_{2}$ and $S_{3}$ in stereographic projection (\ref{confo2}) vanish  and this means that the separable states is projected  to complex plane.
\par
\subsection{Entanglement in $\mathcal{H}_{4}^{\mathbb{C}}\otimes\mathcal{H}_{N}^{\mathbb{C}}$ Hilbert space}
In this section, we  generalize relation between octonionic stereographic projection and concurrence measure for a  bipartite pure states in Hilbert space $\mathcal{H}_{4}^{\mathbb{C}}\otimes\mathcal{H}_{N}^{\mathbb{C}}$. For motivation, let us first consider a pure state $|\psi\rangle\in\mathcal{H}_{4}^{\mathbb{C}}\otimes\mathcal{H}_{N}^{\mathbb{C}}$  in  the following
generic form,
\begin{equation}\label{4part}
|\psi\rangle=\sum\limits_{i = 0}^3 {\sum\limits_{j = 0}^{N-1} {{a_{ij}}\left| {ij} \right\rangle } } ,
\end{equation}
where $|i\rangle (i=0,1, 2, 3)$  and  $|j\rangle (j=0,1,...,N-1)$ are orthonormal real basis of Hilbert space $\mathcal{H}_{4}^{\mathbb{C}}$ and $\mathcal{H}_{N}^{\mathbb{C}}$ respectively. The entanglement of $|\psi\rangle$ is given by concurrence $C$ as
\begin{equation}\label{canc2}
C =2 \sqrt {\sum\limits_{i < j}^3{\sum\limits_{k <l}^{N-1} {|{a_{ik}}{a_{jl}} - {a_{il}}{a_{jk}}{|^2}}} }.
\end{equation}
The state (\ref{4part}) can be written in the octonionic form with the following representation
\begin{align}\label{octo1}
\begin{array}{c}
a_{0j}|0j\rangle \ \ \   \longrightarrow \ \ \  \ \ \ \ \ \ a_{0j}|j\rangle_{o}, \\
a_{1j}|1j\rangle \ \ \   \longrightarrow  \ \ \ \ \ \ a_{1j}e_{2}|j\rangle_{o}, \\
a_{2j}|2j\rangle \ \ \   \longrightarrow \ \ \  \ \ \  a_{2j} e_{4}|j\rangle_{o}, \\
a_{3j}|3j\rangle \ \ \   \longrightarrow \ \ \bar{a}_{3j}e_{2} e_{4}|j\rangle_{o}.
\end{array}
\end{align}
Then the octonionic form of state  (\ref{4part}) can be written as
\begin{equation}\label{4qpar}
|\psi\rangle_{o}= {\sum\limits_{j = 0}^{N-1} {{o_{j}}\left| j \right\rangle_{o} } } ,
\end{equation}
where $o_{j}=(a_{0j}+a_{1j}e_{2})+(a_{2j}+\bar{a}_{3j}e_{2})e_{4}$.
For two octonion
\begin{align}\label{octo3}
\begin{array}{c}
o_k=a_{k0}+a_{k1}e_{2}+(a_{k2}+\bar{a}_{k3}e_{2})e_{4}, \\
o_l=a_{l0}+a_{l1}e_{2}+(a_{l2}+\bar{a}_{l3}e_{2})e_{4},
\end{array}
\end{align}
the stereographic projection is given by
\begin{equation}\label{4qpart}
o_k \bar{o}_l=(S^{0}_{kl}+S^{1}_{kl}e_{2}+(S^{2}_{kl}+S^{3}_{kl}e_{2})e_{4}),
\end{equation}
where
\begin{align}\label{cmot}
\begin{array}{c}
  {S_{kl}^0} =a_{k0}\bar{a}_{l0}+ a_{k1}\bar{a}_{l1}+a_{k2}\bar{a}_{l2}+a_{k3}\bar{a}_{l3},\\
  {S_{kl}^1} =a_{k1}a_{l0} -a_{k0}a_{l1}+\bar{a}_{k3}\bar{a}_{l2}-\bar{a}_{k2}\bar{a}_{l3},\\
 {S_{kl}^2} =a_{k2}a_{l0} -a_{k0}a_{l2}+\bar{a}_{k1}\bar{a}_{l3}-\bar{a}_{k3}\bar{a}_{l1},\\
  {S_{kl}^3} =a_{k2}a_{l1}-a_{k1}a_{l2}+\bar{a}_{k3}\bar{a}_{l0}-\bar{a}_{k0}\bar{a}_{l3},
\end{array}
\end{align}
and $S_{kl}^1 ,S_{kl}^2$ and $S_{kl}^3$ are entanglement sensitive terms related to octonions $o_{k}$ and $o_{l}$, that appear in concurrence measure (\ref{1(23)}) in the form
\begin{align}\label{canc3}
\begin{array}{c}
\mathcal{C}_{k,l}=2 \sqrt {\sum\limits_{k <l}^{N-1} {(|a_{k1}a_{l0} -a_{k0}a_{l1}|^2+|{a}_{k3}{a}_{l2}-{a}_{k2}{a}_{l3}|^2+|a_{k2}a_{l0} -a_{k0}a_{l2}|^2}}\\
\ \ \ \ \ \ \ \ \  \ \ \ \ \ \ \overline{ +|{a}_{k1}{a}_{l3}-{a}_{k3}{a}_{l1}|^2+|a_{k2}a_{l1}-a_{k1}a_{l2}|^2+|{a}_{k3}{a}_{l0}-{a}_{k0}{a}_{l3}|^2)}.
\end{array}
\end{align}
Putting  Eq. (\ref{canc3}) in (\ref{canc2}) yields
\begin{equation}\label{canc5}
C =2 \sqrt {\sum\limits_{k <l}^{N-1} {\mathcal{C}_{k l}^2} } \ .
\end{equation}
To demonstrate the nature of this method, we give two multi-qubit examples in
the following.
\par
First, let us consider the general GHZ-state as Eq. (\ref{ghz}). Octonionic form of this state is given by $|GHZ_{m}\rangle_{o}=(|0^{\otimes m-2}\rangle_{o}-e_{4}|1^{\otimes m-2}\rangle_{o})/\sqrt{2}$ ($o_{0}=1/\sqrt{2}$ and $o_{1}=-e_{4}/\sqrt{2}$). For this GHZ-state, Eqs.  (\ref{canc3}) and (\ref{canc5}) give $C^{(12), \{m-2\}} =1$.
\par
As another example, let us consider the general W-state defined in Eq. (\ref{w}). Octonion form of this state is given by $|W_{m}\rangle_{o}=(|0...001\rangle_{o}+|0...010\rangle_{o}+...+|1...000\rangle_{o})/\sqrt{m}+(e_{2}|00...000\rangle_{o})/\sqrt{m}$ ($o_{0}=...=o_{m-3}=1/\sqrt{m}$ and $o_{m-2}=(e_{2}+e_{4})/\sqrt{m}$). For this state we obtain $C^{(12),\{m-2\}}=\frac{2\sqrt{2(m-2)}}{m}$.
\section{Conclusion}
In conclusion, using the idea of  projecting  one-qubit pure states on usual complex plane, we  theoretically studied the
role of stereographic projection for bipartite pure states in  $\mathcal{H}_{2}^{\mathbb{C}}\otimes \mathcal{H}_{N}^{\mathbb{C}}$ and $\mathcal{H}_{4}^{\mathbb{C}}\otimes \mathcal{H}_{N}^{\mathbb{C}}$  Hilbert spaces.
Since the two-qubit states in complex field are  equivalent to the one-quaterbits in quaternion skew field,  hence  they  could be mapped, using  quaternionic stereographic projection,  into the quaternionic plane. The complex and quaternionic parts of projection are related to the Schmidt term and  concurrence measure  respectively. We found that the  time evolution of any pure two-qubit state, under the general local unitary transformations, is independent of the second unitary operation. Geometrically, this was due the fact that the time evolution is established in  the fiber space of Hopf fibration.  Physically, this   means  that  the local unitary operations do not change the quantum correlation i.e., the concurrence term is invariant under the local unitary transformations.
This result was extended to the bipartite pure states which belong to  Hilbert space $\mathcal{H}_{2}^{\mathbb{C}}\otimes \mathcal{H}_{N}^{\mathbb{C}}$.
\par
In the same manner, we transformed a three-qubit pure state in an octobit, i.e. in a one-qubit form but in octonionic numbers which was generalized to states belong to  the Hilbert space $\mathcal{H}_{4}^{\mathbb{C}}\otimes \mathcal{H}_{N}^{\mathbb{C}}$. The  invariant terms, appearing in generalized concurrence measure, do not changed under local unitary transformations \cite{g.n2}. One may tempt to generalize these trends to the bipartite pure states in Hilbert space $\mathcal{H}_{6}^{\mathbb{C}}\otimes \mathcal{H}_{N}^{\mathbb{C}}$. In this case we have to use  Sednion  numbers  which  is not a division algebra.
\par
The octonionic stereographic projection can be used to classify  $N = 2$ supersymmetric STU
black holes, as it is related to classification of three-qubit entanglements using Fano plane \cite{Duff1}
At the microscopic level, the black holes are described by intersecting $D3$-branes whose wrapping around the six compact dimensions torus $T^6$ provides the string-theoretic interpretation of the charges. Moreover, the Bekenstein-Hawking
black hole entropy \cite{Bekenstein,Hawking} is provided by the three-way entanglement measure via the relation $S=\frac{\pi}{2}\sqrt{\tau_{123}}$ where $S$ is entropy and $\tau_{123}$ is hyperdeterminant \cite{Miyake}.
For a review of the theory of black-hole/qubit correspondence we refer to the article \cite{Duff2}.

\par
\textbf{Acknowledgments}\\
The authors also acknowledge the support from the Mohaghegh Ardabili University.

\end{document}